# A van der Waals Floating-Gate Platform for Broadband Optoelectronic Neuromorphic Computing

Thi Phuong Anh Bach[1], Sangeun Cho[1]*

[1]Division of System Semiconductor, Dongguk University, Seoul, South Korea

## Abstract

The realization of artificial synapses that seamlessly integrate nonvolatile memory with broadband optoelectronic learning is essential for next-generation neuromorphic computing, yet remains challenging because conventional floating-gate memories suffer from limited optical functionality and poor scalability. Here, we report a van der Waals $ReS_2$/hBN/InSe floating-gate synaptic transistor that combines efficient charge-storage memory with broadband optoelectronic learning in a single device architecture. The heterostructure employs InSe as a floating-gate layer for stable charge storage, hexagonal boron nitride (hBN) as a tunneling dielectric, and $ReS_2$ as a photoresponsive channel, enabling efficient carrier tunneling under both electrical and optical stimulation. Benefiting from the synergistic operation of the heterostructure, the device exhibits a large memory window, a high on/off current ratio exceeding $10^6$, stable multilevel conductance states, and reliable retention over $10^4$ s. Furthermore, broadband visible-to-near-infrared optical excitation enables controllable synaptic weight modulation, allowing the emulation of essential synaptic functions, including short- and long-term plasticity, paired-pulse facilitation, pulse-amplitude- and pulse-duration-dependent plasticity, together with optical potentiation and electrical depression. Beyond fundamental synaptic behaviors, the device further demonstrates higher-order learning through associative and discrimination conditioning inspired by the Pavlovian learning paradigm, as well as reconfigurable Boolean logic enabled by electro-optical inputs. These results establish a multifunctional van der Waals floating-gate platform that integrates nonvolatile memory, broadband optoelectronic learning, and adaptive information processing within a single transistor, providing a promising strategy for compact and energy-efficient neuromorphic hardware.

**Keywords:** $ReS_2$, InSe, optoelectronic, artificial synapse, floating gate, neuromorphic

## 1. Introduction

Artificial intelligence (AI) is rapidly transforming modern information technologies, driving an increasing demand for hardware capable of processing and storing massive amounts of information with high speed, low power consumption, and adaptive learning capability. However, conventional computing systems based on the von Neumann architecture fundamentally suffer from the physical separation of memory and processing units, resulting in excessive data transfer, high energy consumption, and limited computational efficiency. In contrast, the human brain seamlessly integrates sensing, memory, computation, and learning through massively interconnected neurons and synapses while operating with remarkable energy efficiency. Inspired by this biological architecture, neuromorphic computing has emerged as a promising paradigm for overcoming the intrinsic limitations of conventional computing systems. Artificial synaptic devices constitute one of the key building blocks for neuromorphic hardware because they emulate biological synaptic plasticity through programmable conductance modulation. Various device platforms, including memristors, phase-change memories, ferroelectric transistors, floating-gate memories, and transistor-based synapses, have been extensively investigated. Nevertheless, many of these technologies still suffer from stochastic switching, filament instability, limited memory windows, low on/off current ratios, and poor retention characteristics, hindering their implementation in large-scale neuromorphic systems [1−5]. Among these approaches, optoelectronic synaptic devices have attracted increasing attention because they combine electrical memory with optical information processing within a single device. Compared with purely electrical synapses, optical stimulation enables remote programming, wavelength-selective operation, and broadband information processing, providing additional degrees of freedom for neuromorphic learning. Incorporating a floating-gate (FG) layer further improves memory retention, multilevel conductance modulation, and long-term synaptic plasticity through stable charge storage [6−9]. However, conventional silicon-based FG memories encounter severe scaling challenges, including charge leakage, short-channel effects, and degraded retention associated with increasingly thin dielectric layers, motivating the exploration of alternative material systems [10−13].

Two-dimensional (2D) van der Waals (vdW) materials provide an attractive solution because of their atomically thin nature, dangling-bond-free interfaces, excellent electrostatic controllability, tunable electronic structures, and strong light-matter interaction. More importantly, arbitrary stacking of different 2D materials enables precise engineering of band alignment and charge-transfer pathways, making vdW heterostructures highly suitable for multifunctional optoelectronic synaptic devices [14–18]. Graphene has been widely employed as a floating-gate material because of its exceptional carrier mobility; however, its zero-bandgap electronic structure inevitably results in a high off-state current and limited current modulation, restricting its applicability to high-performance optoelectronic memory devices [18–26]. In contrast, indium selenide (InSe), a representative III–VI layered semiconductor, possesses high carrier mobility, efficient charge-storage capability, and a thickness-dependent direct bandgap spanning from the ultraviolet to the near-infrared spectral region [27–29]. These characteristics allow InSe to function simultaneously as a stable floating-gate memory layer and an optically active material for broadband neuromorphic operation [30–35]. As an active channel material, rhenium disulfide ($ReS_2$) offers additional advantages owing to its unique distorted crystal structure, which preserves a direct bandgap from bulk to monolayer with weak interlayer coupling. Consequently, $ReS_2$ exhibits stable electrical transport and broadband photoresponse regardless of layer thickness, making it an attractive channel material for optoelectronic synaptic transistors [36]. Integrating $ReS_2$ with hBN and InSe into a vdW heterostructure therefore provides an effective strategy for combining efficient charge storage, broadband optical modulation, and electrically programmable conductance within a single transistor architecture.

Here, we report a van der Waals $ReS_2$/hBN/InSe floating-gate optoelectronic synaptic transistor that integrates nonvolatile memory with broadband optoelectronic learning. In the proposed heterostructure, InSe functions as the floating-gate layer for efficient charge storage, hBN serves as the tunneling dielectric, and $ReS_2$ acts as the photoresponsive channel, enabling dynamic carrier tunneling under both electrical and optical stimulation. The device exhibits a large memory window, a high on/off current ratio exceeding $10^6$, stable multilevel conductance states, and reliable retention over $10^4$ s. Furthermore, broadband visible-to-near-infrared optical stimulation enables diverse synaptic functions, including short- and long-term plasticity (STP/LTP), paired-pulse facilitation (PPF), pulse-amplitude- and pulse-duration-dependent plasticity (PADP/PDDP), together with optical potentiation and electrical depression. Beyond these fundamental synaptic behaviors, the device further demonstrates higher-order learning through associative and discrimination conditioning inspired by the Pavlovian learning paradigm, as well as reconfigurable Boolean logic operations via electro-optical modulation. These results establish a multifunctional vdW floating-gate platform that integrates nonvolatile memory, broadband optoelectronic learning, and adaptive information processing within a single transistor, providing a promising strategy for compact and energy-efficient neuromorphic hardware.

## 2. Results and Discussion

**Figure 1** presents the device concept, heterostructure configuration, and structural characterization of $ReS_2$/hBN/InSe floating-gate synaptic transistor. As illustrated in **Figure 1a**, the device is designed to enable both electrical and optical modulation of the channel conductance within a single van der Waals (vdW) heterostructure. The vertically stacked architecture consists of an InSe floating-gate, an hBN tunneling dielectric, and a $ReS_2$ transport channel on a $SiO_2$/$p^{++}$ Si substrate. Source and drain electrodes are directly contacted with the $ReS_2$ channel, while the highly doped Si substrate serves as the back gate. Electrical gate pulses modulate charge storage in the InSe floating-gate, whereas visible-to-near-infrared (VIS–NIR) optical excitation modulates the postsynaptic current (PSC) through photoinduced carrier generation and charge redistribution within the heterostructure. This architecture therefore provides a common device framework for nonvolatile charge storage and electro-optical synaptic modulation. Details of the device fabrication process are provided in **Figure S1** (Supporting Information). **Figure 1b** shows the crystal structure of the constituent vdW materials. $ReS_2$ possesses a low-symmetry distorted 1T (1T′) structure, which gives rise to anisotropic electronic and optical properties and relatively weak interlayer coupling. Consequently, $ReS_2$ can retain a direct bandgap over a broad thickness range, making multilayer $ReS_2$ attractive for optoelectronic device applications [37].

InSe is a layered III–VI semiconductor with high carrier mobility and thickness-dependent electronic and optical properties, providing favorable characteristics for charge storage and optoelectronic modulation [38]. Meanwhile, hBN provides an atomically flat vdW interface and a wide bandgap, allowing it to serve as an insulating tunneling barrier between the $ReS_2$ channel and the InSe floating-gate [39]. The complementary electronic functions of these three layered materials form the basis of the $ReS_2$/hBN/InSe floating-gate architecture. The fundamental charge-storage mechanism is schematically illustrated in **Figure 1c**. Under an appropriate gate pulse, carriers in the $ReS_2$ channel can be transferred across the hBN tunneling barrier and stored in the InSe floating-gate, resulting in a persistent modulation of the $ReS_2$ channel conductance. Application of an opposite gate polarity reverses the charge-transfer process and releases the stored carriers from the floating-gate. The reversible trapping and detrapping of carriers therefore provides the physical basis for nonvolatile memory operation and electrically programmable synaptic weight modulation. The detailed programming and erasing characteristics and their corresponding energy-band configurations are discussed below.

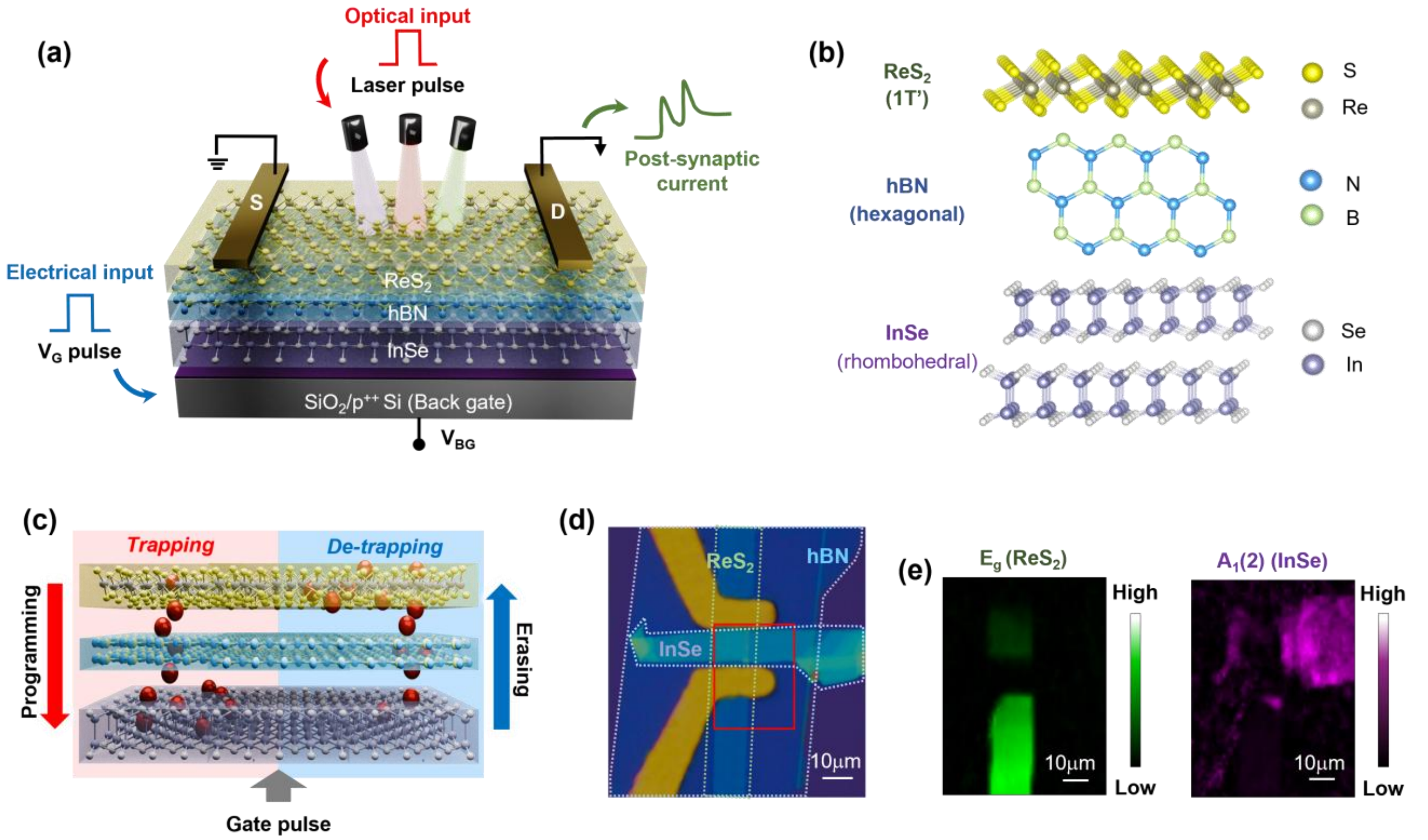

**Figure 1. Device concept and structural characterization of the $ReS_2$/hBN/InSe floating-gate synaptic transistor** (a) Schematic illustration of the vertically stacked $ReS_2$/hBN/InSe floating-gate synaptic transistor, showing electrical gate-pulse and VIS–NIR optical stimulation and the corresponding postsynaptic current (PSC). (b) Crystal structures of $ReS_2$, hBN, and InSe. (c) Schematic illustration of carrier trapping and detrapping across the hBN tunneling barrier during programming and erasing, respectively. (d) Optical microscopy image of the fabricated device, with the boundaries of the constituent flakes indicated by dashed lines. (e) Raman intensity maps of the characteristic $E_g$ mode of $ReS_2$ and $A_1(2)$ mode of InSe acquired from the selected heterostructure region.

The successful assembly of the vertically stacked heterostructure was confirmed by optical and spectroscopic characterization. **Figure 1d** shows an optical microscopy image of the fabricated device, where the boundaries of the $ReS_2$, hBN, and InSe flakes are indicated by the corresponding dashed lines. Atomic force microscopy measurements reveal thicknesses of approximately 4.5, 10, and 13 nm for $ReS_2$, hBN, and InSe, respectively (**Figure S2**, Supporting Information). Raman spectroscopy further confirms the presence of the constituent materials within the heterostructure (**Figure S3**, Supporting Information). The characteristic Raman modes of $ReS_2$ are observed at approximately 149 and 210 $cm^{-1}$, while the peaks at approximately 115, 177, and 225 $cm^{-1}$ are assigned to the characteristic vibrational modes of InSe [40]. The hBN-related mode is additionally observed at approximately ~1364 $cm^{-1}$, confirming the successful integration of all three vdW layers. To spatially verify the heterostructure formation, Raman intensity mapping was performed over the selected device region. Figure 1e displays the Raman maps obtained from the characteristic $E_g$ mode of $ReS_2$ and $A_1(2)$ mode of InSe. The corresponding spatial distributions reproduce the geometries of the individual flakes and clearly identify their overlapping region, confirming the formation of the $ReS_2$/hBN/InSe heterostructure. The $ReS_2$ Raman intensity is reduced in the overlap region, suggesting an interlayer electronic interaction and charge-transfer process between $ReS_2$ and InSe through insulating layer [41]. Photoluminescence measurements further reveal optical transition energies of approximately 1.5 and 1.27 eV for $ReS_2$ and InSe, respectively (**Figure S4**, Supporting Information). Together with the corresponding energy-band alignment shown in **Figure S5** (Supporting Information), these structural and optical characteristics establish the material basis for the charge-storage and broadband optoelectronic behavior discussed in the following sections.

The nonvolatile memory characteristics of the $ReS_2$/hBN/InSe floating-gate synaptic transistor were systematically investigated at room temperature. **Figure 2a** shows the transfer characteristics measured at a constant drain-source voltage ($V_{\mathrm{DS}}$) of 1 V with sweeping the back-gate voltage ($V_{\mathrm{BG}}$) over progressively increasing ranges from ±10 to ±60 V. The device exhibits typical n-type transport with pronounced counterclockwise hysteresis between the forward and reverse gate-voltage sweeps. Notably, the memory window ($\Delta V$) progressively increases as the $V_{\mathrm{BG}}$ sweep range is extended, indicating enhanced charge storage in the InSe floating gate under larger applied gate-voltage. The corresponding programming and erasing processes are schematically illustrated in **Figure 2b**. Under a positive $V_{\mathrm{BG}}$ pulse, the gate-induced electric field drives electrons from the $ReS_2$ channel across the hBN tunneling barrier into the InSe floating gate (**Figure 2b(i)**). The injected electrons remain stored in the electrically isolated InSe layer after removal of the gate pulse, resulting in a persistent modulation of the $ReS_2$ channel conductance and establishing the low-current program state. Conversely, application of a negative $V_{\mathrm{BG}}$ pulse reverses the electric field and removes the stored electrons from the InSe floating gate across the hBN barrier (**Figure 2b(ii)**), thereby restoring the high-current erase state. These reversible programming and erasing processes, governed by charge storage and removal in the InSe floating gate, provide the physical basis for the nonvolatile memory operation of the device.

To distinguish the floating-gate effect from other possible contributions to the observed hysteresis, a series of control measurements were performed. The output characteristics measured at different $V_{\mathrm{BG}}$ values exhibit nearly linear $I_{\mathrm{DS}} - V_{\mathrm{DS}}$ behavior (**Figure S6**, Supporting Information), indicating approximately ohmic contacts between the metal electrodes and the $ReS_2$ channel. In addition, control $ReS_2$ and InSe field-effect transistors fabricated separately on $SiO_2$/p$^{++}$-Si substrates exhibit n-type and p-type transport characteristics, respectively (**Figure S7**, Supporting Information). In particular, the control $ReS_2$ transistor exhibits negligible hysteresis under comparable gate-sweep conditions.

These observations support that the pronounced hysteresis of the $ReS_2$/hBN/InSe device predominantly originates from charge storage in the InSe floating gate rather than from the intrinsic hysteresis of the $ReS_2$ channel or contact-related effects.

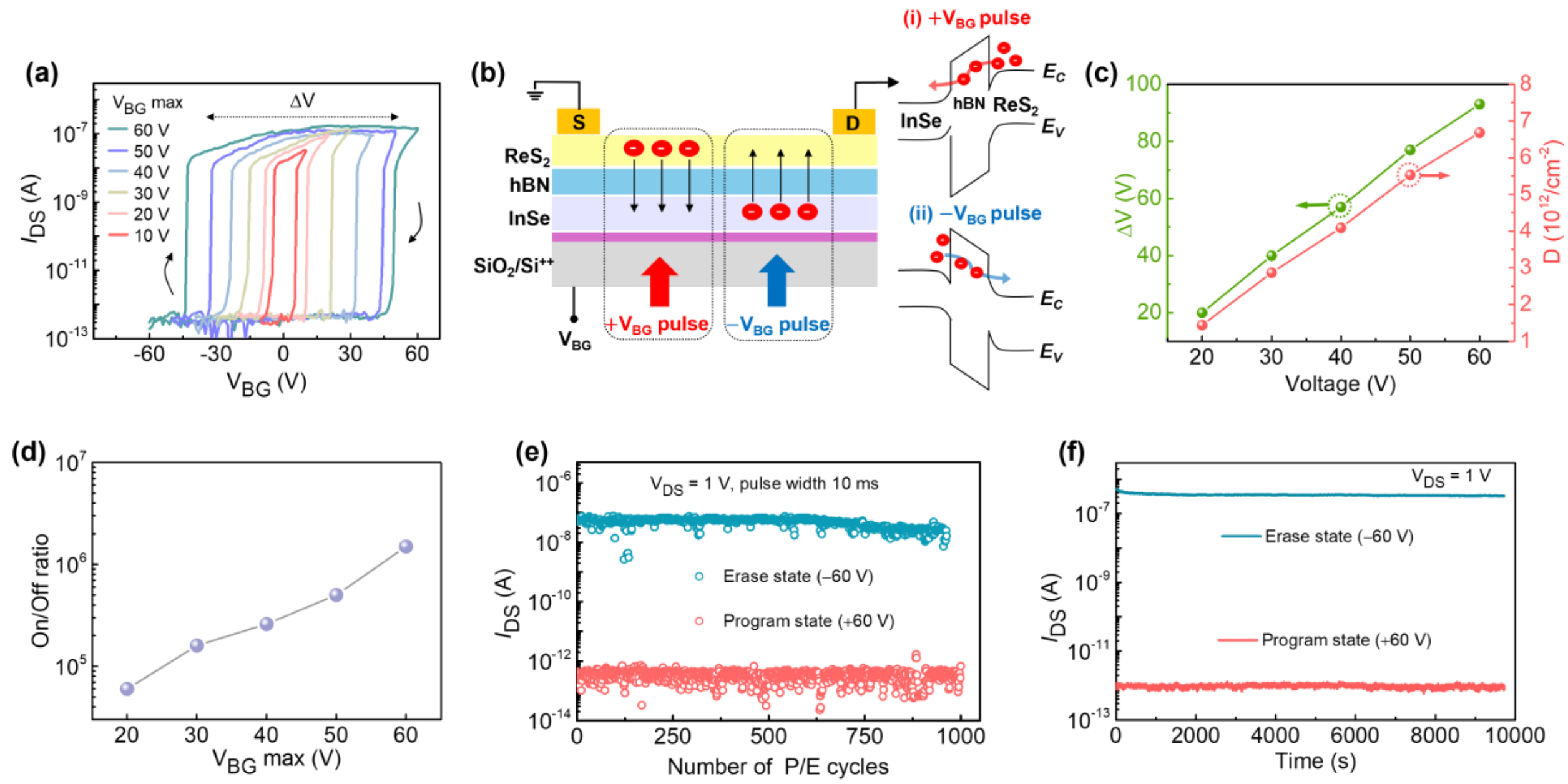

**Figure 2. Nonvolatile memory characteristics of the $ReS_2$/hBN/InSe floating-gate transistor** (a) Transfer characteristics measured at $V_{DS}$=1 V for different back-gate voltage sweep ranges from ±10 to ±60 V. (b) Schematic illustration of the floating-gate memory operation under positive and negative $V_{BG}$ pulses, showing charge storage during programming and charge release during erasing. (c) Memory window ($\Delta V$) and extracted stored charge density ($D$) as a function of the gate-voltage sweep range. (d) On/off current ratio as a function of the gate-voltage sweep range. (e) Endurance characteristics over 1000 program/erase cycles using ±60 V pulses with a duration of 10 ms. (f) Retention characteristics of the program and erase states measured for approximately $10^4$ s.

The charge-storage capability of the device was quantitatively evaluated using the $\Delta V$ defined as the difference between the threshold voltages extracted from the forward and reverse transfer curves. As shown in **Figure 2c**, $\Delta V$ increases approximately linearly as the $V_{BG}$ sweep range is extended and reaches ~ 93 V for the ±60 V sweep. The ratio of the memory window to the total gate-voltage sweep range correspondingly approaches ~80% at ± 60 V (**Figure S8**, Supporting Information), demonstrating pronounced nonvolatile threshold-voltage modulation.

To demonstrate the superior charge-trapping capability of the InSe FG, the charge density stored in the FG was calculated using the following equation [42]:

$$D = \frac{(\Delta V \times C_{CG-FG})}{e} \quad (1)$$

where $e$ is the electron charge and $C_{CG-FG}$ is the capacitance per unit area between the control gate and the FG. $C_{CG-FG}$ was calculated using $C_{CG-FG} = \frac{\varepsilon_0 \times \varepsilon_r}{d}$, where $\varepsilon_0$, $\varepsilon_r$, and $d$ are the vacuum permittivity, relative dielectric constant, and thickness of the $SiO_2$ dielectric layer, respectively.

As shown by the red data in **Figure 2c**, the estimated stored charge density increases with the $V_{BG}$ sweep range, following the evolution of the memory window, and reaches $6.67 \times 10^{12}$ $cm^{-2}$ at a sweep range of ±60 V. The large stored charge density is consistent with the pronounced memory window and demonstrates the effective charge-storage capability of the InSe floating gate.

The magnitude of the nonvolatile conductance modulation also increases with the $V_{BG}$ sweep range. As shown in **Figure 2d**, the on/off current ratio increases monotonically with increasing sweep range and exceeds $10^6$ at a gate-voltage sweep limit of $\pm 60$ V The corresponding $I_{DS} - V_{DS}$ characteristics measured after programming and erasing with $+60$ V and $-60$ V gate pulses, respectively, with a duration of 10 ms, exhibit clearly distinguishable conductance states with an on/off ratio exceeding $10^6$ (**Figure S9**, Supporting Information). The influence of the heterostructure geometry on the memory characteristics was further investigated by varying the thicknesses of the hBN tunneling dielectric and the InSe floating-gate layer. In particular, the memory window exhibits a pronounced dependence on the hBN thickness and reaches its largest value among the investigated devices at an hBN thickness of approximately 10 nm (**Figure S10**, Supporting Information), highlighting the importance of the tunneling-barrier thickness in determining the charge-storage characteristics. A corresponding dependence on the InSe floating-gate thickness is presented in **Figure S11** (Supporting Information), further demonstrating the critical role of the heterostructure geometry in controlling the memory performance.

The operational reliability of the device was further evaluated through endurance and retention measurements. As shown in **Figure 2e**, repeated application of $+60$ V programming and $-60$ V erasing pulses with a duration of 10 ms reproducibly switches the device between the low-current program state and the high-current erase state over 1000 program/erase (P/E) cycles. The two conductance states remain clearly distinguishable throughout the cycling test without substantial degradation. **Figure 2f** further demonstrates the nonvolatile retention characteristics of the device. Following programming and erasing with $+60$ V and $-60$ V pulses, respectively, both conductance states remain well separated for approximately $10^4$ s without additional gate stimulation. Collectively, the large memory window, high on/off ratio, reproducible endurance, and stable retention demonstrate robust nonvolatile memory operation of the $ReS_2$/hBN/InSe floating-gate transistor, providing the basis for the programmable synaptic functionalities discussed below.

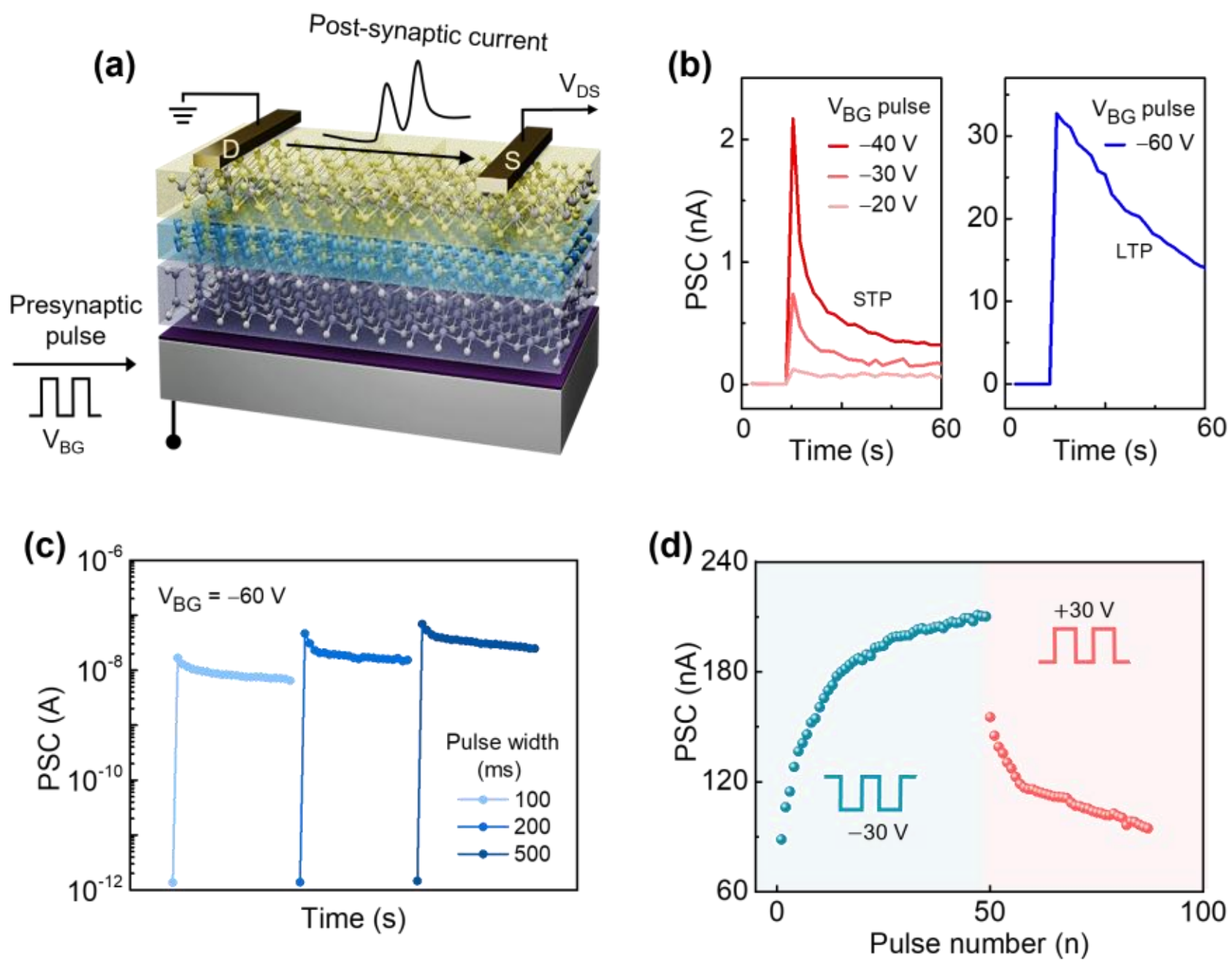

**Figure 3. Electrically driven synaptic plasticity of the $ReS_2$/hBN/InSe floating-gate transistor** (a) Schematic illustration of electrical synaptic operation, in which back-gate voltage pulses serve as presynaptic stimuli and the drain current is monitored as the postsynaptic current (PSC). (b) Temporal PSC responses to negative $V_{BG}$ pulses with different amplitudes, demonstrating the evolution from short-term plasticity (STP) toward long-term plasticity (LTP). (c) PSC responses to −60 V pulses with different pulse widths of 100, 200, and 500 ms, demonstrating pulse-duration-dependent plasticity. (d) Gradual potentiation and depression of the PSC induced by consecutive −30 and +30 V electrical pulses, respectively.

Following the demonstration of nonvolatile memory operation, we investigated the electrically driven synaptic plasticity of the $ReS_2$/hBN/InSe floating-gate transistor. As schematically illustrated in **Figure 3a**, voltage pulses applied through the $V_{BG}$ serve as presynaptic stimuli, while the resulting drain current is monitored as the postsynaptic current (PSC). This configuration enables the synaptic weight of the device to be modulated by controlling the amplitude, duration, polarity, and number of electrical pulses.

**Figure 3b** shows the temporal PSC responses to negative $V_{BG}$ pulses with different amplitudes. At relatively low pulse amplitudes (−20 to −40 V), the PSC increases immediately upon stimulation and subsequently decays toward its initial state, resembling short-term plasticity (STP). Increasing the pulse amplitude leads to a larger PSC response and a progressively slower decay after removal of the stimulus. At −60 V, a substantial fraction of the enhanced PSC persists over the measurement period, indicating a transition toward long-term plasticity (LTP). The extracted change in PSC increases monotonically with increasing magnitude of the negative $V_{BG}$ pulse (**Figure S12**, Supporting Information), confirming pulse-amplitude-dependent plasticity (PADP). This behavior demonstrates that the strength of the electrically induced synaptic response can be systematically tuned by the amplitude of the presynaptic stimulus.

Synaptic plasticity can also be modulated by the duration of the electrical stimulus. **Figure 3c** shows the PSC responses to $-60$ V pulses with widths of 100, 200, and 500 ms. Increasing the pulse width produces a larger conductance modulation and a more persistent PSC response, demonstrating pulse-duration-dependent plasticity (PDDP). Together with the amplitude-dependent behavior, these results indicate that the synaptic weight can be continuously regulated by both the strength and duration of the electrical stimulus. Conversely, positive $V_{\mathrm{BG}}$ pulses induce an inhibitory response, with the PSC progressively decreasing as the pulse amplitude increases (**Figure S13**, Supporting Information). The opposite conductance modulation induced by negative and positive gate pulses therefore enables bidirectional control of the synaptic weight. The capability for multilevel conductance modulation was further examined using individual negative $V_{\mathrm{BG}}$ pulses ranging from $-20$ to $-60$ V in 5 V increments with a fixed pulse width of 10 ms. As shown in **Figure S14** (Supporting Information), different pulse amplitudes produce distinguishable conductance states, demonstrating electrically programmable multilevel memory behavior. Such multilevel programmability allows multiple synaptic weights to be represented within a single device and provides the basis for analog-like weight modulation beyond binary memory operation.

Finally, the cumulative modulation of the synaptic weight was investigated through consecutive potentiation and depression pulses. As shown in **Figure 3d**, repeated $-30$ V pulses progressively increase the PSC, corresponding to synaptic potentiation, whereas subsequent $+30$ V pulses gradually decrease the PSC, corresponding to synaptic depression. The gradual and bidirectional evolution of the conductance under consecutive electrical stimuli emulates the long-term potentiation (LTP) and long-term depression (LTD) processes of biological synapses. These results demonstrate controllable and cumulative electrical modulation of the synaptic weight in the $ReS_2$/hBN/InSe floating-gate transistor.

Beyond electrically driven synaptic plasticity, the $ReS_2$/hBN/InSe floating-gate transistor also enables direct optical modulation of the synaptic state. In biological visual systems, incident light is converted into electrical signals by photoreceptors and subsequently processed through synaptic networks. Inspired by this light-to-electrical signal transduction, the $ReS_2$/hBN/InSe heterostructure was investigated as an optoelectronic synapse in which optical stimuli directly modulate the postsynaptic current.

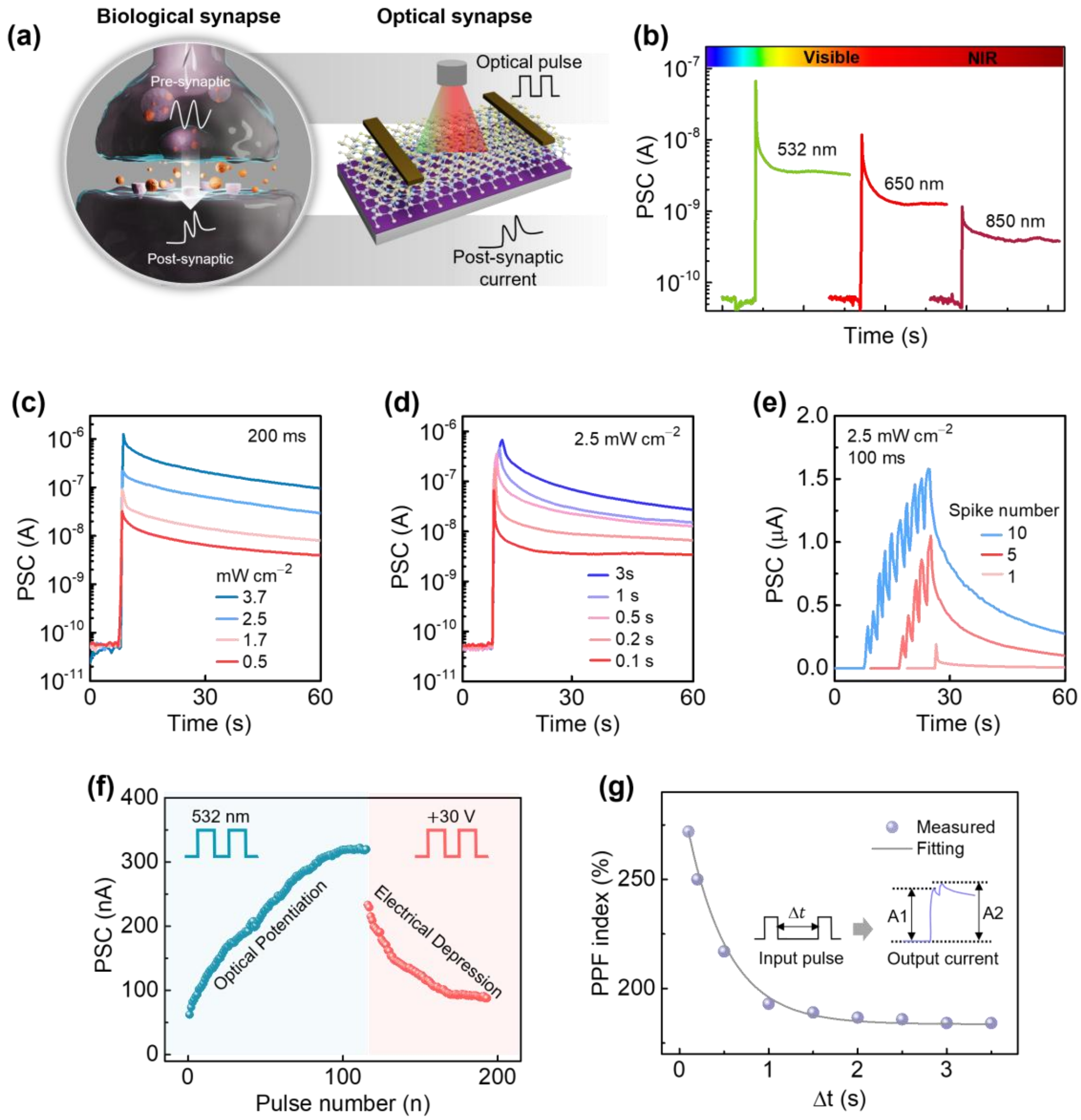

**Figure 4. Broadband optoelectronic synaptic plasticity of the $ReS_2$/hBN/InSe floating-gate transistor** (a) Schematic comparison between a biological synapse and the optoelectronic synaptic transistor, in which optical pulses serve as presynaptic stimuli. (b) Temporal PSC responses under 532, 650, and 850 nm illumination, demonstrating optoelectronic synaptic operation across the VIS–NIR range. (c) PSC responses to 532 nm illumination with different optical power densities at a fixed illumination duration of 200 ms. (d) PSC responses under different illumination durations at a fixed optical power density of 2.5 mW cm$^{-2}$. (e) Cumulative PSC modulation as a function of the number of consecutive optical pulses. (f) Optical potentiation induced by consecutive 532 nm pulses followed by electrical depression induced by +30 V gate pulses. (g) Paired-pulse facilitation (PPF) index as a function of the inter-pulse interval ($\Delta t$)under 532 nm illumination. The inset illustrates the definition of $A_1$, $A_2$ , and $\Delta t$.

As schematically illustrated in **Figure 4a**, optical pulses serve as presynaptic stimuli, while the resulting drain current is monitored as the PSC. A key feature of the device is its optical operation over a broad spectral range extending from the visible (VIS) to near-infrared (NIR) region**. Figure 4b** shows the temporal PSC responses under illumination at 532, 650, and 850 nm with a constant optical power density of 1.7 mW cm$^{-2}$. Distinct PSC responses are observed at all three wavelengths, demonstrating optoelectronic synaptic operation over the investigated wavelength range of 532−850 nm. Notably, the device retains a measurable and persistent PSC response even under 850 nm excitation, extending its optical synaptic functionality into the NIR region. The magnitude of the PSC exhibits a clear wavelength dependence, reflecting the wavelength-dependent photoresponse of the $ReS_2$/hBN/InSe heterostructure. Under visible illumination at 532 and 650 nm, photoexcitation generates a pronounced conductance modulation, whereas the response decreases under 850 nm illumination. The corresponding energy-band configurations under VIS and NIR excitation are presented in **Figure S15** (Supporting Information). Among the investigated wavelengths, 532 nm illumination produces the largest PSC modulation and was therefore selected for further investigation of optically induced synaptic plasticity.

The optical synaptic weight can be systematically regulated by the characteristics of the incident light. **Figure 4c** shows the PSC responses under 532 nm illumination with power densities ranging from 0.5 to 3.7 mW $cm^{-2}$ at a fixed illumination duration of 200 ms. Increasing the optical power density progressively enhances the PSC and produces a more persistent conductance state after removal of the optical stimulus. Similarly, increasing the illumination duration from 0.1 to 3 s at a fixed power density of 2.5 mW $cm^{-2}$ results in progressively enhanced and prolonged PSC responses (**Figure 4d**). The synaptic response is also dependent on the number of optical stimuli. As shown in **Figure 4e**, increasing the number of consecutive optical pulses from 1 to 10 produces a cumulative enhancement of the PSC. The tunability of the optical synaptic response was further investigated under 650 nm illumination, as shown in **Figure S16** (Supporting Information). These power-, duration-, and pulse-number-dependent responses demonstrate that the optically induced synaptic weight can be continuously modulated through multiple parameters of the optical stimulus. Under relatively weak or short illumination, the induced conductance modulation relaxes comparatively rapidly, whereas stronger or longer optical stimulation produces enhanced and increasingly persistent conductance states, consistent with a gradual evolution from short-term toward long-term plasticity. The corresponding changes in PSC ($\Delta PSC$), extracted as a function of optical power density and illumination duration, are summarized in **Figure S17** (Supporting Information), further demonstrating the systematic tunability of the optically induced synaptic response. Importantly, the device further supports complementary optical and electrical modulation of the synaptic weight. As shown in **Figure 4f**, successive 532 nm optical pulses with a power density of 2.5 mW $cm^{-2}$ and a duration of 0.2 s progressively increase the PSC, corresponding to optical potentiation. Subsequent +30 V electrical pulses with the same duration gradually decrease the PSC, producing electrical depression. Thus, the conductance state can be increased by optical stimulation and decreased independently by electrical stimulation, providing complementary optoelectronic control of the synaptic weight. This capability enables optical information to be directly encoded into the device conductance while retaining an independent electrical pathway for regulating the stored synaptic state, making the device particularly attractive for sensory neuromorphic systems.

Paired-pulse facilitation (PPF), a representative form of short-term synaptic plasticity, was further examined using two consecutive optical pulses separated by an inter-pulse interval ($\Delta t$). The PPF index was calculated as

$$PPF\ (\%) = \frac{A_2}{A_1} \times 100 \qquad (2)$$

where $A_1$ and $A_2$ re the amplitudes of the first and second PSC responses, respectively. **Figure 4g** reveals the dependence of the PPF index on $\Delta t$ under 532 nm illumination. At a short interval of 0.1 s, the device exhibits a PPF index of ~270%, indicating pronounced facilitation of the second PSC response. As $\Delta t$ increases, the PPF index gradually decreases and approaches a nearly saturated value, reflecting the progressive relaxation of the transient state generated by the first optical stimulus before arrival of the second pulse. This interval-dependent facilitation reproduces the temporal correlation characteristic of biological short-term synaptic plasticity and further demonstrates the capability of the device to encode the temporal relationship between consecutive optical stimuli. Building on these tunable optical synaptic characteristics, the device further exhibits experience-dependent learning behavior. As shown in **Figure S18** (Supporting Information), consecutive 532 nm optical pulses progressively increase the PSC during the initial learning process, followed by spontaneous conductance relaxation after removal of the optical stimulus, representing forgetting. While 11 optical pulses are required to reach a predetermined PSC level during the initial learning stage, only 7 pulses are required to recover the same level during relearning after partial relaxation. The reduced number of stimuli required during relearning demonstrates a history-dependent synaptic response, providing a basis for the higher-order learning behaviors discussed below.

Discrimination learning represents a more selective form of associative learning in which a system learns to respond differently to distinct stimuli according to their reinforcement history. As a fundamental form of associative learning, Pavlovian classical conditioning was first reproduced in our device, as detailed in **Figure S19**, Supporting Information. Inspired by this behavior, we implemented a discrimination-conditioning paradigm using two electrical stimuli with different amplitudes together with an optical unconditioned stimulus (US). As schematically illustrated in **Figure 5a**, the experiment consists of three sequential phases: before, during, and after conditioning.

Electrical pulses of −10 V in 2 s were designated as the unreinforced conditioned stimulus (CS−), representing the low-pitched bell, whereas −15 V pulses with the same duration were designated as the reinforced conditioned stimulus (CS+), representing the high-pitched bell. Optical pulses of 532 nm with a power density of 0.5 mW $cm^{-2}$ and a duration of 0.2 s served as the US, analogous to food in the Pavlovian paradigm. A PSC of 1 nA was defined as the response threshold for distinguishing an effective response.

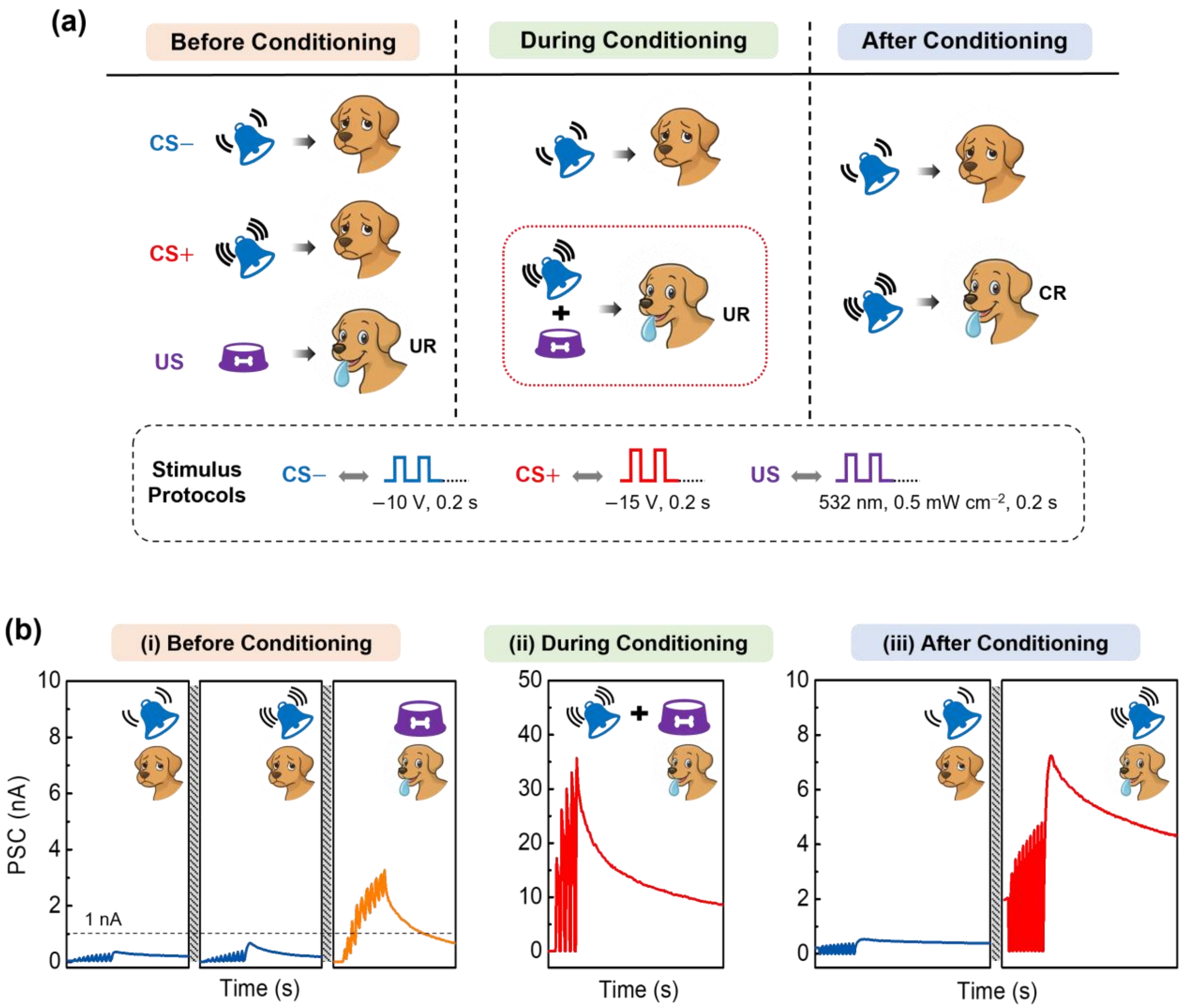

**Figure 5. Discrimination conditioning in the $ReS_2$/hBN/InSe optoelectronic synaptic transistor** (a) Schematic illustration of the discrimination-conditioning paradigm consisting of before-, during-, and after-conditioning stages. Two electrical stimuli are assigned as the unreinforced conditioned stimulus (CS−) and reinforced conditioned stimulus (CS+), respectively, while 532 nm optical pulses serve as the unconditioned stimulus (US) that intrinsically evokes the unconditioned response (UR). During conditioning, CS+ is repeatedly paired with the optical US, whereas CS− is presented without reinforcement.(b) Experimental PSC responses during (i) the before-conditioning stage, where CS− and CS + alone produce responses below the predefined 1 nA threshold, whereas the optical US evokes an UR above the threshold; (ii) the conditioning stage, where CS+ is repeatedly paired with the optical US while CS− remains unreinforced; and (iii) the after-conditioning stage, where CS+ alone evokes a conditioned response (CR) above the threshold, whereas the response to CS− remains below the threshold.

Before conditioning, 10 electrical pulses of −10 V and −15 V individually produced only weak responses below the predefined 1 nA threshold, whereas 10 optical pulses generated a PSC exceeding the threshold, corresponding to the unconditioned response (UR) (**Figure 5b(i)**). Thus, prior to associative training, neither CS− nor CS+ alone was sufficient to evoke the salivation-like response, whereas the optical US readily elicited the designated response. This initial state establishes that neither electrical cue alone is sufficient to evoke the designated response before associative training. During conditioning, CS+ is repeatedly paired with the optical US. As shown in **Figure 5b(ii)**, repeated CS+–US pairing produces a pronounced enhancement of the PSC, reaching approximately 36 nA. The discriminative response becomes evident after conditioning. As shown in **Figure 5b(iii)**, presentation of CS+ alone produces a PSC of approximately 7 nA, clearly exceeding the 1 nA response threshold, whereas the response to CS− alone remains below the threshold. Thus, CS+ evokes a conditioned response (CR) even in the absence of the optical US, while the unreinforced CS− fails to evoke the corresponding response. Importantly, the observed discrimination cannot be attributed solely to the difference in electrical pulse amplitude, because both CS− and CS produce responses below the predefined threshold before conditioning. Instead, the selective post-conditioning response reflects their different reinforcement histories. These results demonstrate stimulus-selective associative learning in a single $ReS_2$/hBN/InSe floating-gate synaptic transistor, extending the device functionality beyond basic Pavlovian association toward discrimination conditioning.

Beyond associative and discrimination learning, the electro-optical responsiveness of the $ReS_2$/hBN/InSe floating-gate transistor was further exploited to implement reconfigurable Boolean logic operations. The ability to process electrical and optical signals within the same device provides an additional functionality for integrating sensing, memory, and information processing in a compact optoelectronic architecture. In particular, the VIS–NIR optical response demonstrated above enables optical signals of different wavelengths to directly participate in logic operations. As schematically illustrated in **Figure 6a**, the electrical gate pulse ($E_{\text{in}}$) and optical pulse ($O_{\text{in}}$) are defined as two independent logic inputs, while the resulting drain current serves as the logic output ($E_{\text{out}}$). The presence and absence of each input stimulus are assigned as logic “1” and “0”, respectively. By adjusting the electrical and optical stimulation conditions relative to a predefined output-current threshold, the same device can be configured to perform either AND or OR logic operations. The corresponding input-output relationships are summarized in **Figure 6b**.

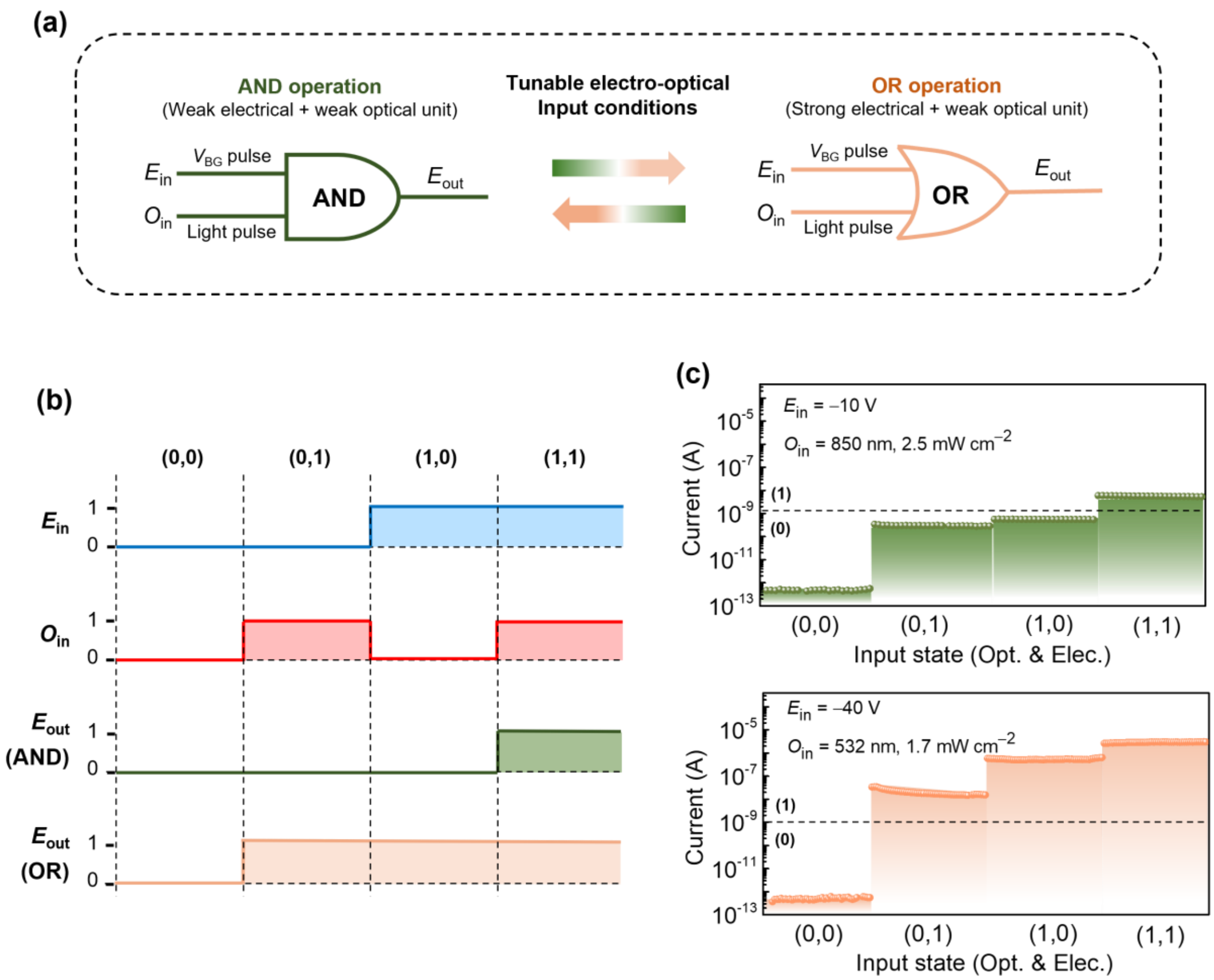

**Figure 6. Reconfigurable electro-optical Boolean logic operations in the $ReS_2$/hBN/InSe floating-gate transistor** (a) (Schematic illustration of the reconfigurable AND and OR logic operations using electrical ($E_{in}$) and optical ($O_{in}$) stimuli as two independent inputs and the drain current as the output ($E_{out}$). (b) Input and output logic states for the four input combinations, (0,0), (0,1), (1,0), and (1,1), corresponding to AND and OR operations. (c) Experimentally measured output currents for the four input states. The AND operation is implemented using a −10 V electrical input and 850 nm optical input (2.5 mW cm$^{-2}$), whereas the OR operation is realized using a −40 V electrical input and 532 nm optical input (1.7 mW cm$^{-2}$). The dashed horizontal lines indicate the predefined current thresholds separating logic "0" and "1".

For the AND operation, a −10 V electrical pulse and 850 nm optical illumination with a power density of 2.5 mW cm$^{-2}$ were used as $E_{in}$ and $O_{in}$, respectively. As shown in the upper panel of **Figure 6c**, the output current remains below the predefined threshold when no stimulus is applied (0,0) or when either the optical (0,1) or electrical (1,0) input is applied individually. In contrast, simultaneous application of both inputs (1,1) increases the output current above the threshold, yielding logic "1". The corresponding output sequence therefore follows the AND truth table: 0, 0, 0, and 1 for the input states (0,0), (0,1), (1,0), and (1,1), respectively. The device can be reconfigured to perform OR logic by changing the stimulation conditions. For this operation, a −40 V electrical pulse and 532 nm optical illumination with a power density of 1.7 mW cm$^{-2}$ were employed. As shown in the lower panel of **Figure 6c**, either the optical input (0,1) or the electrical input (1,0) alone generates an output current above the threshold, while simultaneous application of both stimuli (1,1) also maintains the high-output state. Only the absence of both inputs (0,0) produces logic "0", resulting in the OR output sequence of 0, 1, 1, and 1. These results demonstrate that the logic functionality of the $ReS_2$/hBN/InSe transistor can be reconfigured between AND and OR operations by tuning the electrical and optical input conditions without changing the physical device architecture. Notably, the AND operation under 850 nm illumination demonstrates that NIR optical signals can directly participate in Boolean information processing. Together with the nonvolatile memory, electrically and optically tunable synaptic plasticity, and discrimination learning demonstrated above, the reconfigurable electro-optical logic functionality highlights the versatility of the $ReS_2$/hBN/InSe floating-gate transistor for integrated neuromorphic information processing.

Finally, the overall performance of the $ReS_2$/hBN/InSe floating-gate synaptic transistor was benchmarked against representative previously reported floating-gate synaptic devices, as summarized in **Table 1** (Supporting Information). The comparison considers key memory and synaptic characteristics, including the memory-window-to-sweep-range ratio, stored charge density, on/off ratio, retention, PPF index, available stimulation modes, and optical operating wavelength. The present device exhibits a memory-window-to-sweep-range ratio of 78%, a stored charge density of $6.67 \times 10^{12}$ $cm^{-2}$, an on/off ratio exceeding $10^6$, retention over $10^4$ s, and a PPF index of ~272%. More importantly, the device combines electrical synaptic modulation with broadband VIS–NIR optical operation extending from 532 to 850 nm within a single floating-gate architecture. Together with the associative and discrimination learning demonstrated above, these characteristics highlight the integrated memory and optoelectronic learning capabilities of the $ReS_2$/hBN/InSe floating-gate transistor.

## 3. Conclusion

In summary, we demonstrated a multifunctional $ReS_2$/hBN/InSe van der Waals floating-gate synaptic transistor that integrates nonvolatile memory, electrically and optically tunable synaptic plasticity, higher-order learning, and reconfigurable logic within a single heterostructure. The InSe floating gate enables effective charge storage through the hBN tunneling barrier, resulting in a large memory window, an on/off current ratio exceeding $10^6$, stable retention over ~$10^4$ s, and reliable endurance over 1000 program/erase cycles. Beyond conventional memory operation, the device exhibits tunable electrical synaptic plasticity and broadband optoelectronic modulation across the VIS–NIR range from 532 to 850 nm, with the synaptic weight systematically controlled by the amplitude, duration, and number of electrical and optical stimuli. The device further reproduces experience-dependent and higher-order learning behaviors, including learning–forgetting–relearning, associative learning, and discrimination conditioning, demonstrating both history-dependent plasticity and stimulus-selective responses. Moreover, complementary optical potentiation and electrical depression enable electro-optical control of the synaptic weight, while the same device realizes reconfigurable AND and OR Boolean logic operations by adjusting the electrical and optical input conditions. The integration of nonvolatile memory, broadband optoelectronic plasticity, selective learning, and reconfigurable information processing within a single van der Waals floating-gate transistor establishes a versatile device architecture for compact optoelectronic neuromorphic hardware.

## 4. Experimental Section

*Device Fabrication*: Bulk $ReS_2$ and InSe crystals were commercially obtained from 2D Semiconductors (USA), while bulk hBN crystals were purchased from HQ Graphene. Few-layer $ReS_2$, hBN, and InSe flakes were mechanically exfoliated from the corresponding bulk crystals using adhesive tape. The exfoliated flakes were sequentially assembled on a 300 nm $SiO_2/p^{++}$-Si substrate using a dry-transfer technique assisted by a polydimethylsiloxane (PDMS) stamp. The InSe flake was first transferred onto the $SiO_2/p^{++}$-Si substrate to serve as the floating-gate layer, followed by hBN as the tunneling dielectric and $ReS_2$ as the transport channel, forming the vertically stacked $ReS_2$/hBN/InSe van der Waals heterostructure. A schematic illustration of the device fabrication process is provided in **Figure S1** (Supporting Information). Source and drain electrodes were subsequently defined on the $ReS_2$ channel by conventional photolithography, followed by thermal evaporation of Cr/Au (5/50 nm) under $\sim 2\times10^{-6}$ Torr. The fabrication process was completed by lift-off in acetone.

*Material and Structural Characterization*: The morphology and thicknesses of the exfoliated $ReS_2$, hBN, and InSe flakes were characterized by atomic force microscopy (AFM; Bruker MultiMode 8-HR). Raman spectroscopy, Raman mapping and Photoluminescence (PL) were performed using a JASCO NRS-4500 Raman spectrometer with a 532 nm excitation laser to identify the constituent materials, spatially verify the formation of the heterostructure and investigate the optical characteristics of $ReS_2$ and InSe. All structural and optical characterizations were performed at room temperature under ambient condition.

*Electrical and Synaptic Measurements*: All electrical measurements, including transfer curves, $I-V$ characteristics and synaptic measurements, were carried out using a Keithley 4200-SCS semiconductor parameter analyzer connected to a probe station under ambient conditions at room temperature. Unless otherwise specified, the drain-source voltage was fixed at 1 V during the transfer and synaptic measurements. The nonvolatile memory characteristics were evaluated by measuring the drain current ($I_{\mathrm{DS}}$) while sweeping the back-gate voltage ($V_{\mathrm{BG}}$) over different voltage ranges. Program/erase operations and electrically induced synaptic responses were generated using an Agilent 81110A pulse/pattern generator. The resulting drain current was monitored as the postsynaptic current (PSC). For the endurance measurements, program and erase pulses of $+60$ V and $-60$ V, respectively, with a duration of 10 ms were repeatedly applied, and the corresponding conductance states were read at $V_{\mathrm{DS}}$=1 V. For the retention measurements, the program and erase states were monitored at $V_{\mathrm{DS}}$=1 V without additional programming stimuli for approximately $10^4$ s.

*Optoelectronic and Neuromorphic Measurements*: Optoelectronic measurements were performed using laser sources with wavelengths of 532, 650, and 850 nm to evaluate the optical synaptic response over the visible-to-near-infrared (VIS–NIR) range. The incident optical power was controlled using neutral-density filters and calibrated using Newport 1936-R optical power meter. Optically induced synaptic plasticity was characterized by varying the illumination wavelength, optical power density, pulse duration, pulse number, and inter-pulse interval while monitoring the PSC. Paired-pulse facilitation (PPF) measurements were performed using two consecutive optical pulses separated by a controlled time interval ($\Delta t$). For electro-optical potentiation/depression measurements, optical pulses were employed to increase the synaptic conductance, followed by electrical gate pulses to decrease the conductance. Learning–forgetting–relearning and associative-learning measurements were implemented using predefined sequences of electrical and optical pulses, with the detailed stimulation conditions provided in the corresponding figures and captions. Reconfigurable Boolean logic operations were evaluated by defining the electrical gate pulse and optical pulse as two independent logic inputs and determining the logic output from the measured drain-current level relative to a predefined current threshold. Unless otherwise specified, all electrical and optoelectronic measurements were performed under ambient conditions at room temperature.

**Acknowledgements**

This research has been supported by the POSCO Science Fellowship of POSCO TJ Park Foundation.

## Author Contributions

**Thi Phuong Anh Bach:** Methodology, Investigation, Validation, Writing manuscript, **Sangeun Cho**: Writing –review & editing, Resources, Supervision, Project administration.

## Conflict of Interest

The authors declare no conflict of interest.

## Data Availability Statement

The data supporting the findings of this study are available upon reasonable request from the corresponding author.